\documentclass{PoS}
\usepackage{epsfig}
\PoS{PoS(BDMH2004)060} 

\title{The Formation of Galactic Bulges}

\ShortTitle{The Formation of Galactic Bulges}

\author{\speaker{Reynier Peletier}\\
        Kapteyn Institute, Groningen, The Netherlands\\
        E-mail: \email{peletier@astro.rug.nl}} 

\author{Marc Balcells \\
	Instituto de Astrof\'\i sica de Canarias, La Laguna, Spain \\
	E-mail: \email{balcells@ll.iac.es}}
	
\author{Jes\'us Falc\'on-Barroso \\
        Sterrewacht Leiden, Leiden, The Netherlands \\
	E-mail: \email{jfalcon@strw.leidenuniv.nl}}

\author{Alister Graham \\
	Research School of Astronomy and Astrophysics, Mount Stromlo Observatory, Australia  \\
	E-mail: \email{graham@mso.anu.edu.au}} 

\abstract{We summarise some recent results about nearby galactic bulges that are
relevant to their formation. We highlight a number of significant advances 
in our understanding of the surface brightness profiles, stellar populations, 
and especially the very centers of spiral galaxies. We also view our own Milky Way as if
it were an external galaxy. Our main conclusions are that bulges of early-type
spirals (S0 -- Sb) contain central nuclear components, just like late-type
spirals and most other types of galaxies. The luminosities of these central components
correlate best with total bulge luminosity, and not as well with morphological
type. Bulges of early-type spiral galaxies follow the fundamental plane and the 
colour/line strength vs. luminosity relations of elliptical galaxies.
Although we have a reasonable idea about bulges of
early-type spirals we know much less about late-type bulges. However,
the close resemblance of our Milky Way Bulge to bulges in external disk galaxies makes
us suspect that bulges of late-type spirals might be very similar as
well.}

\FullConference{Baryons in Dark Matter Halos\\
		 5-9 October 2004\\
		 Novigrad, Croatia}

\begin{document}

\section{Introduction}

\vspace{-0.2cm}

Galactic bulges --- objects in the
middle of spirals that are similar in many ways to elliptical galaxies, 
are the key to study galaxy formation. Although bulges of
nearby disk galaxies have been studied in great detail in recent years (see Kormendy
\& Kennicutt 2004 for an in-depth review article), many aspects are
still unknown. In this short review we will try to show that, contrary to 15
years ago, we are able to make a fairly coherent picture of the nature of
galactic bulges, which we can use to test formation models. Rather than trying
to comment on all the issues in a much too short space we will concentrate on a
few pieces in this puzzle. Before doing this, however, we will first discuss the
definition of the bulge. After having talked about surface brightness profiles,
stellar populations, and the centers of galaxies, we will end with a short
section about our knowledge of bulges at higher redshift.

\vspace{-0.1cm}

\section{Definition of Bulges}

\vspace{-0.2cm}

One could ask the question whether it makes sense to talk about galactic bulges
as separate components of spiral galaxies. If galaxy properties change smoothly
from the center outwards, as is the case for elliptical galaxies, it is
not clear whether discussing the inner parts separately makes sense. In the
case of bulges, however, one can make the case that they are clearly different from their
surrounding disk. They contain less interstellar matter, are generally rounder,
have a surface brightness profile that is more strongly peaked than the inward extrapolation of the
outer exponential profile, and have
different internal kinematics than the underlying disk. 
In the past, Hubble (1936) noted that {\sl the
normal spiral exhibits a bright, semi-stellar nucleus and a relatively large
nuclear region of unresolved nebulosity which closely resembles a lenticular
(E7) nebula.}  

There are a few methods in the literature that are used to calculate the extent
of the bulge. The first is one-dimensional. Here an exponential disk model
and an r$^{1/4}$ bulge model are fitted to a major axis surface brightness
profile (e.g. Freeman 1970, Kormendy 1977). A generalization of this method 
takes a S\'ersic r$^{1/n}$ profile for the bulge (e.g. Andredakis et al. 1995, de Jong 1996). This 
method can be used for galaxies at all inclinations, but fails when the slope of
the surface brightness profiles in bulge and disk region is the same. A
different method, using the surface brightness distribution together with the
morphology of the bulge, was introduced by Kent (1985). Here a galaxy is
decomposed into a bulge and a disk of fixed, but different ellipticity, using a
non-parametric method. This method uses the fact that in projection bulges are often rounder
than their disks, so that the ellipticity in the central regions
is lower. This method, only applicable for galaxies that are not face-on,
generally gives similar results to the previous one, except in the case of
galaxies with boxy bulges (e.g. Jarvis 1986), i.e., edge-on galaxies with
rounder, generally very boxy, isophotes. One can think about a third method,
using only the stellar kinematics, which decomposes a galaxy into a fast
rotating, cold disk and a slowly rotating, hot bulge. This method, which to our
knowledge has not been applied to real data yet, might be the most physical,
since it separates a galaxy into a slowly rotating elliptical-like spheroid and
a fast-rotating disk. Since kinematic analysis has shown that boxy {\it bulges}
generally are bars seen edge-on (e.g. Bureau \& Freeman 1999), and bars are a
phenomenon that occurs in flat disks, the first method is more in agreement
with the third, and for that reason we will use that for the rest of this
proceedings. 

Apart from decomposing the surface brightness profile in the radial direction
it is also of interest to know how the profile in the vertical direction has to
be decomposed. We will come to this issue in the next Section.

\vspace{-0.1cm}

\section{Surface Brightness Profiles and Components in Bulges}

\vspace{-0.2cm}

\begin{figure}

\vspace{-0.8cm}
\epsfig{figure=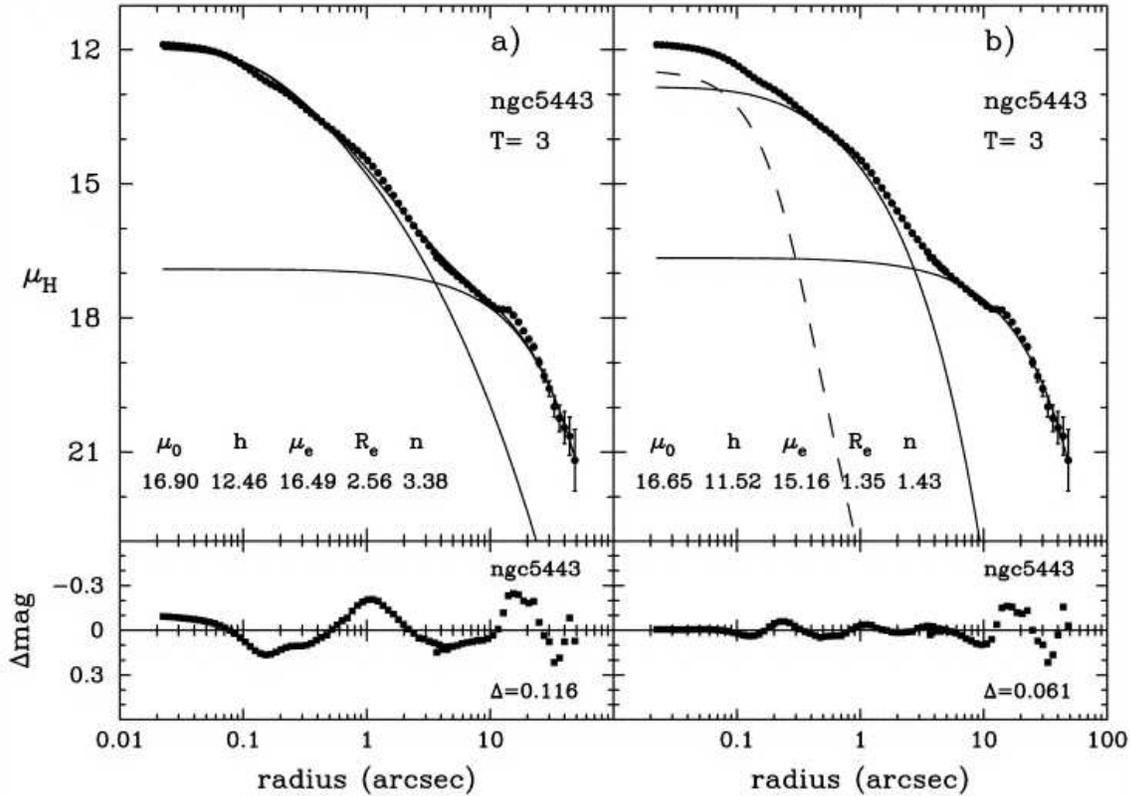, width=1.\textwidth}

\vspace{-0.5cm}

\caption{(a) Combined HST plus ground-based H-band surface brightness profile
of NGC 5443 (Sb), with PSF-convolved S\'ersic bulge and exponential disk
components (solid lines). The residuals (data minus model) are shown in the
lower panel. (b) Same surface brightness profile fitted with a PSF-convolved
S\'ersic bulge, an exponential disk, plus a central point source (dashed line). 
Residuals are shown in the lower panel. From Balcells et al. 2003.}
\label{fig1}
\end{figure}

Inspired by work on elliptical galaxies (Caon et al. 1993) Andredakis et al.
(1995) found that surface brightness profiles of bulges are better fit by a
S\'ersic r$^{1/n}$ law than by the traditional r$^{1/4}$ distribution. With the
aid of higher resolution Hubble Space Telescope photometry, Balcells et al.
(2003) performed the decomposition again on a subsample of the original dataset
of Andredakis et al. The results here are quite interesting: in 84\% of the
sample a better fit is obtained if a central point source is included. In a
moderate fraction of the sample, another increase in the goodness of the fit is
obtained if an extended component, such as a nuclear disk, is included as well.  In
Fig.\ref{fig1} we show figure 1 of Balcells et al. (2003), which shows the
residuals with and without a central point source for an example galaxy.

\begin{figure}

\vspace{-0.8cm}
\epsfig{figure=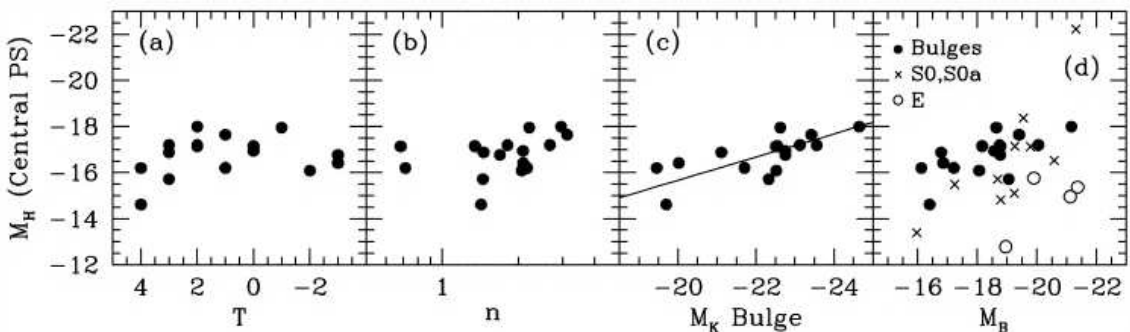, width=1.\textwidth}

\vspace{-0.5cm}

\caption{Dependence of the central point source H--band absolute magnitude M$_{H,PS}$ on 
(a) galaxy morphological
type index T. (b)  S\'ersic index $n$ of the bulge. (c) H-band absolute bulge magnitude. The line is an orthogonal regression to the
distribution. (d) absolute B-band magnitude. Filled circles: Our
bulges; the absolute magnitude is that of the bulge. Crosses: S0 galaxies from
Ravindranath et al. (2001). Open circles: E galaxies from Ravindranath et al.
(2001). Total galaxy absolute magnitudes are used for S0 and E galaxies. From
Balcells et al. (2003)}
\label{fig2}
\end{figure}

Given that it is now commonly accepted that most large galaxies have central
black holes in their centers (e.g. Magorrian et al. 1998, Ferrarese \& Merritt 2000) it is not
surprising that bulges contain central point sources, also since many galaxies
contain stronger or weaker AGNs. In fact, HST studies of elliptical and S0
galaxies have found many central point sources (Ravindranath et al. 2001, Rest et al. 2001), 
in late-type spirals Carollo et al. (2002) detected nuclear sources in 43
of their 69 objects, and in dE's Graham \& Guzman (2003) detected many point sources. 
Interestingly enough, Balcells et al. show that the
nuclear point source luminosity correlates with several quantities, most of all
the luminosity of the bulge (Fig.\ref{fig2}) . The nuclear point sources,
containing up to a fraction of order 10$^{-3}$ of the bulge luminosity,  are
easy to accommodate within either a merger origin or a secular evolution origin
for bulges. The observed scaling of PS and bulge luminosities  does not
necessarily indicate an internal origin for the PS since the nuclear gas
deposition during a merger may scale with the masses of the merging objects. If
bulges are evolved bars, the problem of growing a nuclear star cluster is
similar to that of feeding an active galactic nucleus (Maciejewski et al.
2002). If black hole mass correlates with bulge mass, bulge mass correlates with
luminosity and nuclear cluster luminosity correlates with black hole mass, this
is what one expects. 

\begin{figure}

\vspace{-0.8cm}
\epsfig{figure=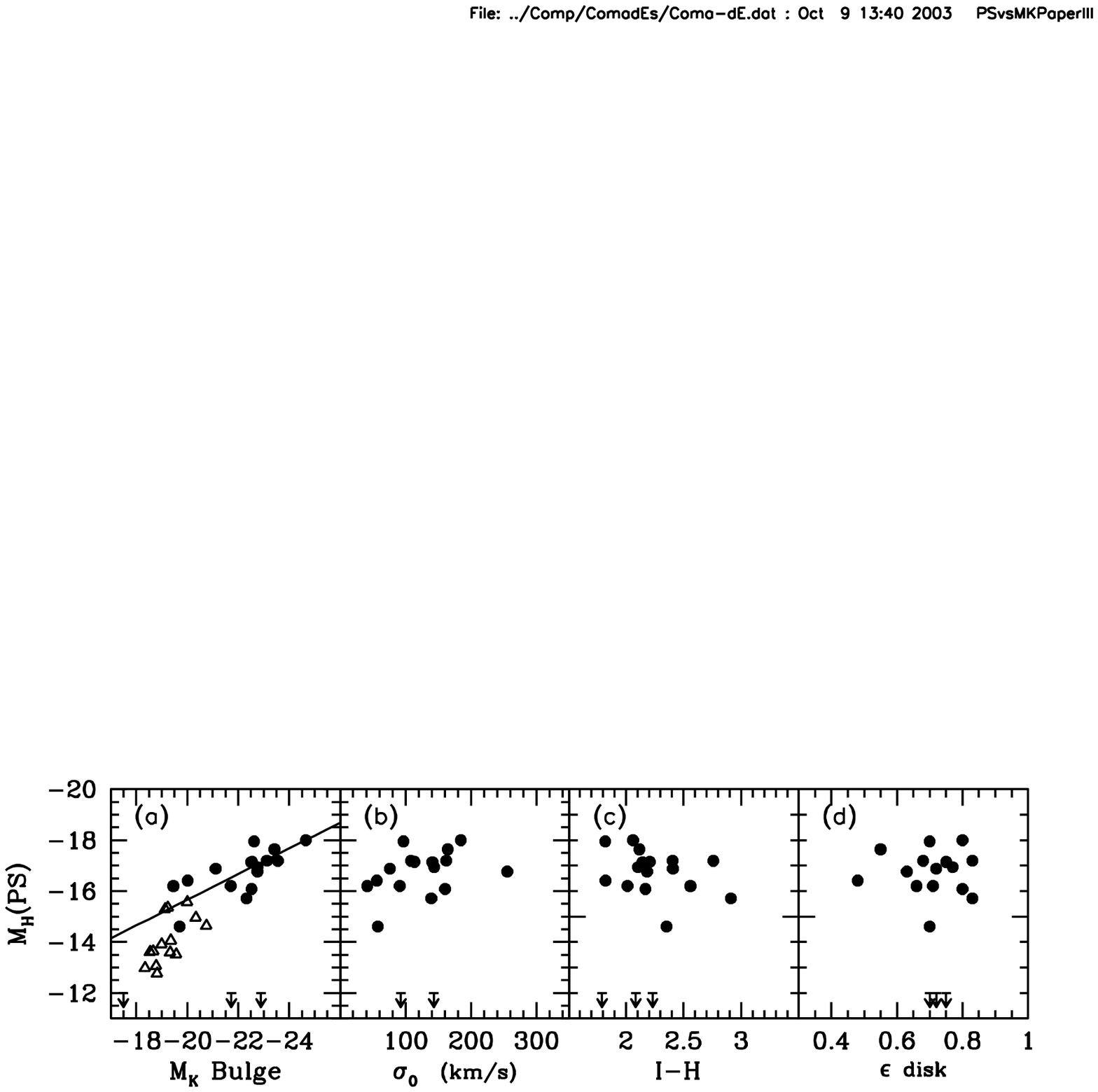, width=1.\textwidth}

\vspace{-0.5cm}

\caption{Absolute $H$--band magnitude of the unresolved nuclear sources as a 
function of: ({\it a}) the $K$--band absolute magnitude of the bulge.  {\it
Filled circles}: Our bulges.  {\it Triangles}: dwarf ellipticals from Graham \&
Guzm\'an (2003). The line is the regression to the bulge data ({\it b})
Aperture-corrected central velocity dispersion; ({\it c}) Central $I-H$ color,
from HST/NICMOS F160W and WFPC2 F814W images (Peletier et al.\ 1999).  ({\it d}) 
Ellipticity of the outer disk, from two-dimensional bulge-disk decomposition in
$K$--band images (Andredakis et al. 1995). Upper limits are given for the three
galaxies without detected point sources; one of these does not have a central
velocity dispersion measurement. From Balcells et al. (2004).}
\label{fig3}
\end{figure}

In Figs.\ref{fig2} and \ref{fig3} we also show some other correlations with the
luminosity of the central point source. The correlation with the S\'ersic index
$n$ is strong, while the correlation with morphological type T is much weaker.
In fact, Balcells et al. (2004) show that all correlations with
morphological type are less strong than with bulge luminosity. 
One of the main conclusions from that paper is that large
galaxies are not scale-free, as Courteau et al. (1996) claim, but that they have
a scale, namely the luminosity of the bulge. Fig.\ref{fig3} also shows that the
correlation with the central velocity dispersion, and the dust-free colour $I-H$
are not as strong as with bulge luminosity, and that there is no correlation
with inclination (disk axis ratio). We would expect a strong correlation with 
central potential well depth, and possibly the correlation of point source
luminosity with velocity dispersion would become stronger if we would make a
correction for rotation. We are currently analysing some Integral Field 
Spectroscopy of this sample to investigate this issue.

Not only from the surface brightness profiles it is apparent that bulges contain
other components, such as nuclear disks, bars, etc., but detailed internal
kinematics also shows this. For example, the minor axis kinematics of this sample, presented in 
Falc\'on-Barroso et al. (2003a), shows structures that are different from major
axis disks. Moreover, SAURON-data of the S0 galaxy  NGC 7332 (Falc\'on-Barroso
et al. 2004) shows that this galaxy most likely has a central disk, a  bar, and a very small kinematically decoupled stellar disk. 
The gas kinematics in this galaxy are completely decoupled from the stellar
kinematics, strongly indicating interactions with the neighbouring galaxy NGC
7339.

\vspace{-0.1cm}

\section{Stellar Populations of Bulges}

\vspace{-0.2cm}

At the moment we know relatively little about the stellar populations of bulges, and
most of this information is limited to bulges of early-type spiral galaxies. This has to
do with the fact that many bulges are dusty, making the interpretation of broadband
colours difficult, and that one has to separate a bulge from an often dominating disk.
For late-type bulges (type Sbc and later) most of our information comes from HST studies
(e.g., B\"oker et al. 2002, 2003; Carollo et al. 2002). We know that late-type
spirals have nuclear clusters that are more prominent than the ones in early-type
spirals. Colours indicate that the central regions are full of dust and young
stars. Given the fact that these bulges are very small compared to
their disks, it is hard to establish whether these young stars and dust are situated in
the disk or in the bulge, we think that it is too early to make definitive statement
about bulge populations. In our own Milky Way one is able to de-contaminate
the bulge from the disk using e.g. proper motions. In this way Zoccali et al. (2003) find
that the age of the Galactic Bulge is old, around 12 -- 13 Gyr.

\begin{figure}
\begin{center}
\vspace{-0.8cm}
\epsfig{figure=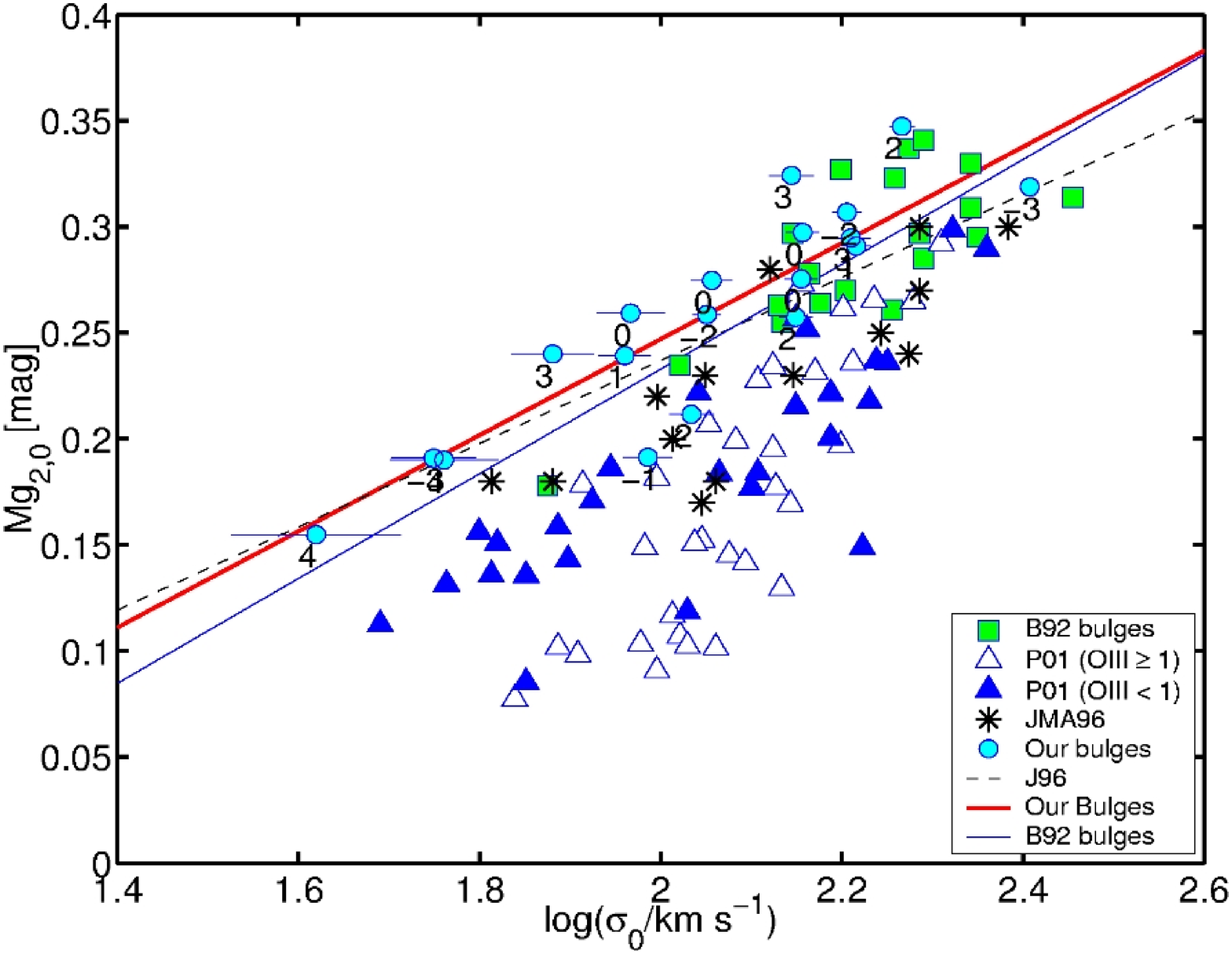, width=\textwidth}
\end{center}

\vspace{-0.5cm}

\caption{The Mg$_{2}$ -- $\sigma$ relation. Filled circles correspond 
to the sample of bulges of Falc\'on-Barroso et al. (2002). Solid squares 
represent the sample of bulges from Bender et al. (1992). 
Morphological T-types of our sample are indicated.
Open triangles are bulges with EW(OIII)$\geq$1, while filled triangles are bulges with 
EW(OIII)$<$1, both from Prugniel et al. (2001). Asteriscs correspond to
the sample of bulges of Jablonka et al. (1996). 
Error bars of log($\sigma_0$) for our sample are also plotted. From
Falc\'on-Barroso et al. (2002).}
\label{fig4}
\end{figure}

For early-type bulges a few significant results have been reported.  Broadband colours
indicate that  bulges are old (10 Gyr with a range of about 2 Gyr), except for the very
central regions (Peletier et al. 1999). The position of bulges on the fundamental plane
of elliptical galaxies is consistent with this statement. Their formation epoch must
have been at most 2.5 Gyr later than elliptical and S0 galaxies in clusters
(Falc\'on-Barroso et al.\ 2002). The situation with absorption line strengths is less
clear. Falc\'on-Barroso et al. (2003b) find that the CaT* -- $\sigma$ relation for
bulges is the same as for elliptical galaxies, but with less scatter, and that CaT* is decreasing
with increasing velocity dispersion, contrary to e.g. the
Mg$_2$ index, which is larger with increasing $\sigma$. It is not clear yet what the
reason is for this decrease. Possibilities are a Ca under-abundance w.r.t. Fe, a
bottom-heavy IMF, or uncertainties in our knowledge of element-enrichment in galaxies
(see also Cenarro et al.\ 2003). The fact that the relation for ellipticals is the same
as for early-type spirals indicates that dust extinction cannot play a major role. The Mg$_2$ --
$\sigma$ relation also shows that  bulges of early-type spirals behave in a
similar way as elliptical galaxies (see Fig.\ref{fig4}). However, although the data of
Falc\'on-Barroso et al. agree with e.g., Jablonka, Martin \& Arimoto (1996) and Bender,
Burstein \& Faber (1992), the agreement with the dataset of Prugniel,
Maubon \& Simien (2001) is not very good. The latter data indicate that, in
general, bulges are much younger than ellipticals. This  difference might have
been caused by a different selection procedure. In the case of the data of
Falc\'on-Barroso et al., much care was taken to make sure that the amount of disk
light contaminating the data was limited. In the case of Prugniel et al.,
however, bulges were selected at random. As a result, it is very likely that
star formation in the disk, on the same line of sight as the bulge light, could
affect the points and cause the difference. However, one should also realise
that the colours of the bulge and inner disk, at least for early type bulges,
are generally very similar (Terndrup et al. 1994, Peletier \& Balcells 1996) and
indicate that the central regions are dominated by old stellar populations. 
Better quality data will have to settle this issue.

\vspace{-0.1cm}

\section{The Centers of Disk Galaxies}

\vspace{-0.2cm}

\begin{figure}
\begin{center}
\vspace{-0.8cm}
\epsfig{figure=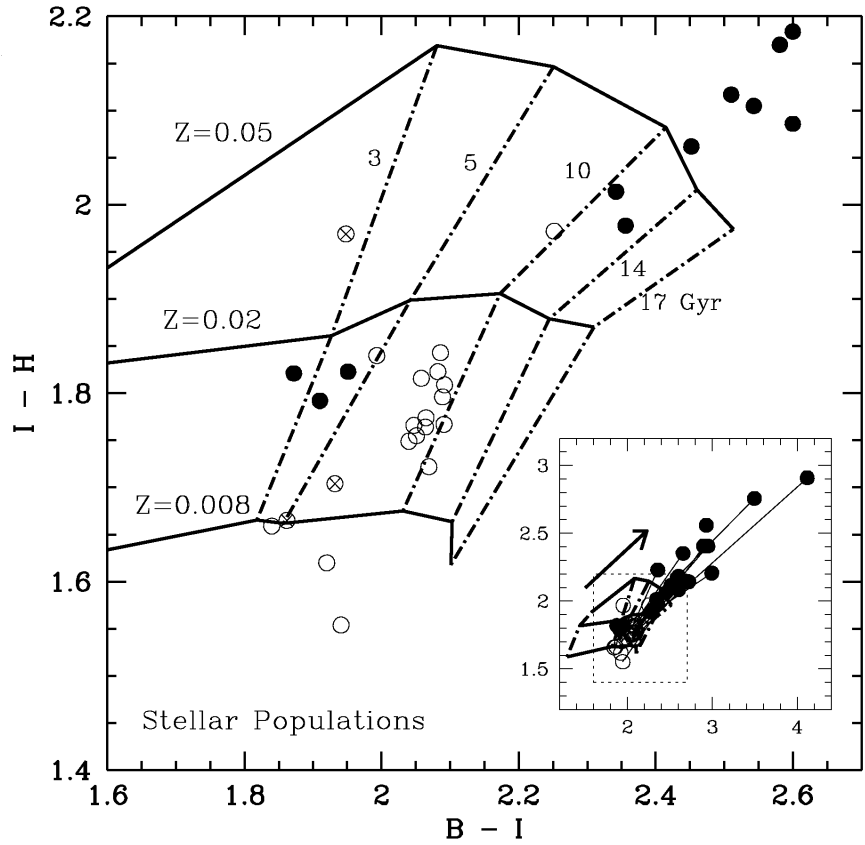, width=0.8\textwidth}
\end{center}

\vspace{-0.5cm}

\caption{Colour-colour diagrams for the 20 galaxies. 
Displayed are the positions of the center (filled) and at 1 bulge effective
radius on the minor axis (open circles). A reddening vector for a reddening of
$A_V$ =  1 mag (inset) is given as well. Sbc galaxies are indicated with
circled crosses. Superimposed are SSP models of Vazdekis et al.\ (1996).
Solid lines are lines of constant metallicity, dashed-dotted lines are loci of
constant age. From Peletier et al.\ (1999)}
\label{fig5}
\end{figure}

In the previous discussion we have not talked about the very central regions of 
spiral galaxies, apart from the central point sources. In Fig.~\ref{fig5} we
have reproduced Fig.~3a from Peletier et al. (1999). It is clear from this figure 
that the filled points for most galaxies are situated far above the model grid,
indicating that they contain dust: the central extinction A$_V$ lies between 0.6 and 1
mag. Very few galaxies do not have such a dusty core. As can be seen clearly in
the color map of NGC~5838, this dust is often associated with central star
formation. So, {\sl although most of the bulge is old, some stars are currently
formed in the very central regions}. In elliptical galaxies a similar phenomenon
is seen, since at least 80\% of them contain dust in their central regions (van
Dokkum \& Franx 1995).  When one looks at absorption line strengths, one also
often sees that the innermost point is younger than the rest of the galaxy
(e.g., Davies et al. 2001; S\'anchez-Bl\'azquez et al. 2004). This is also seen 
in long slit spectroscopic studies of bulges (Jablonka et al. 2002). Since the
material in the center is generally metal rich, the origin of the gas is
probably internal, and not gas infalling from outside.

Drawing the analogy with our Galaxy (type Sbc), we see a similar structure. 
Launhardt et al. (2002), using DIRBE data, find an almost exponential bulge,
with an enhancement in the region of the Galactic Plane (see also Kent et al. 1991). 
In that region, with a
diameter of 250 pc, comparable to the nuclear regions in other galaxies, most of
the light comes from cool stars, with a torus of interstellar matter in the same
region. Outside this torus there is an absolute lack of interstellar matter. In
the very inner parts (r $<$ 1 pc) the light is dominated by hot, massive stars.
This shows that the Milky Way bulge is similar in almost all aspects to external
bulges of galaxies. It shows that even in an Sbc bulge the large majority of
stars in the bulge are old. Careful studies of other late-type spirals will have
to be made to see whether the same is the case there.

\vspace{-0.1cm}

\section{Bulges at High Redshift}

\vspace{-0.2cm}

\begin{figure}
\begin{center}
\vspace{-0.8cm}
\epsfig{figure=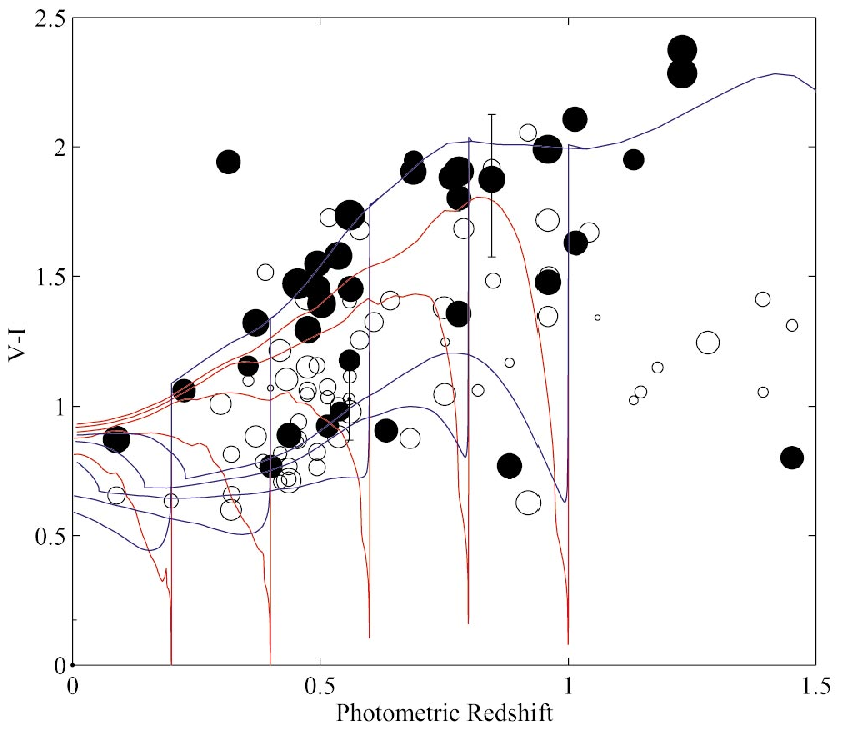, width=0.8\textwidth}
\end{center}

\vspace{-0.5cm}

\caption{V--I colors for ellipticals (solid symbols, large aperture) and spiral bulges
(open symbols, 5\% aperture ) as a function of photometric redshift. Symbol sizes are
proportional to galactic central concentration. The solid blue curve at the top
represents the observed V--I color expected for a passively evolving system that formed
in a single burst at z = 3. This is the baseline model upon which secondary bursts are
added. The other curves define color-redshift trajectories for systems that suffer
secondary bursts of activity at redshifts of 0.2,0.4,0.6,0.8, and 1. Red curves refer to
0.1 Gyr bursts involving 15\% of the stellar mass, whereas the blue curves refer to
extended bursts with 5 Gyr e-folding time scales. From Ellis et al. (2001).} 
\label{fig6}
\end{figure}

Fig.\ref{fig6} is a diagram from Ellis et al.\ (2001) showing the colour 
evolution as a function of redshift for galaxies in the Hubble Deep Field. Since
here the open symbols are bulges and the filled symbols ellipticals, one would
think that at a given redshift the bulges are bluer, and therefore younger or
more metal poor. This conclusion, unfortunately, cannot be drawn from this
diagram alone. Bulges are small (bulges that dominate the disk across a radial
range of more than 1 kpc are rare), so even in the HDF their size is small. We
will have to wait for studies that take into account the contribution from the
disk, when studying the bulge. These studies might come soon, i.e. from studies
like GOODS, and will lead to interesting progress in the field. 

Although we haven't discussed the formation of bulges themselves here, all the
results presented here are relevant to obtain the right idea about how spirals
formed. For this topic we refer to the review of Kormendy \& Kennicutt (2004) 
and references in there.

\vspace{-0.3cm}

\begin{thebibliography}{99}

\vspace{-0.3cm}


\bibitem{And95} Andredakis, Y.~C., 
Peletier, R.~F., \& Balcells, M.\ 1995, MNRAS, 275, 874 
\bibitem{Bal03} Balcells, M., Graham, 
A.~W., Dom{\'{\i}}nguez-Palmero, L., \& Peletier, R.~F.\ 2003, ApJl, 582, 
L79 
\bibitem{Bal04} Balcells, M., Graham, A. \& Peletier, R.F., 2004, astro-ph/0404379
\bibitem{Ben92} Bender, R., Burstein, 
D., \& Faber, S.~M.\ 1992, ApJ, 399, 462
\bibitem{Bok03} B{\" o}ker, T., 
Stanek, R., \& van der Marel, R.~P.\ 2003, AJ, 125, 1073 
\bibitem{Bok02} B{\" o}ker, T., 
Laine, S., van der Marel, R.~P., Sarzi, M., Rix, H., Ho, L.~C., \& Shields, 
J.~C.\ 2002, AJ, 123, 1389 
\bibitem{Bur99} Bureau, M., \& 
Freeman, K.~C.\ 1999, AJ, 118, 126 
\bibitem{Cao93} Caon, N., Capaccioli, M., 
\& D'Onofrio, M.\ 1993, MNRAS, 265, 1013
\bibitem{Car02} Carollo, C.~M., 
Stiavelli, M., Seigar, M., de Zeeuw, P.~T., \& Dejonghe, H.\ 2002, AJ, 
123, 159 
\bibitem{Cen03} Cenarro, A.~J., Gorgas, 
J., Vazdekis, A., Cardiel, N., \& Peletier, R.~F.\ 2003, MNRAS, 339, L12 
\bibitem{Cou96} Courteau, S., de Jong, 
R.~S., \& Broeils, A.~H.\ 1996, ApJl, 457, L73 
\bibitem{Dav01} Davies, R.~L., et al.\ 
2001, ApJl, 548, L33 
\bibitem{deJ96} de Jong, R.~S.\ 1996, A\&AS, 
118, 557 
\bibitem{Ell01} Ellis, R.~S., Abraham, 
R.~G., \& Dickinson, M.\ 2001, ApJ, 551, 111 
\bibitem{Fal02} Falc{\'o}n-Barroso, J., Peletier, R.~F., Balcells, M.\ 2002,
MNRAS, 335, 741.
\bibitem{Fal03a} Falc{\'o}n-Barroso, J., Balcells, M., Peletier, R.~F., \&
Vazdekis, A.\ 2003, A\&A, 405, 455 
\bibitem{Fal03b} Falc{\'o}n-Barroso, J.,  Peletier, R.~F., Vazdekis, A. \&
Balcells, M.,\ 2003, ApJ, 588, L17.
\bibitem{Fal04} Falc{\' 
o}n-Barroso, J., et al.\ 2004, MNRAS, 350, 35 
\bibitem{Fer00} Ferrarese, L., \& 
Merritt, D.\ 2000, ApJl, 539, L9 
\bibitem{Fre70} Freeman, K.~C.\ 1970, ApJ, 
160, 811 
\bibitem{Gra03} Graham, A.~W., 
\& Guzm{\' a}n, R.\ 2003, AJ, 125, 2936 
\bibitem{Hub36} Hubble, E.~P.\ 1936, Yale 
University Press,  
\bibitem{Jab96} Jablonka, P., Martin, 
P., \& Arimoto, N.\ 1996, AJ, 112, 1415 
\bibitem{Jab02} Jablonka, P., Gorgas, 
J., \& Goudfrooij, P.\ 2002, ApSS, 281, 367 
\bibitem{Jar86} Jarvis, B.~J.\ 1986, AJ, 91, 
65
\bibitem{Ken85} Kent, S.~M.\ 1985, ApJs, 59, 115
\bibitem{Ken91} Kent, S., Dame, T. \& Fazio, G.G., 1991, ApJ, 378, 131
\bibitem{Kor77} Kormendy, J.\ 1977, ApJ, 
218, 333 
\bibitem{Kor04} Kormendy, J., \& 
Kennicutt, R.~C.\ 2004, ARAA, 42, 603 
\bibitem{Lau02} Launhardt, R., Zylka, 
R., \& Mezger, P.~G.\ 2002, A\&A, 384, 112
\bibitem{Mac02} Maciejewski, W.\ 2002, 
Astronomical Society of the Pacific Conference Series, 275, 251 
\bibitem{Mag98} Magorrian, J. et al., 1998, AJ, 115, 2285
\bibitem{Pel96} Peletier, R.~F., 
\& Balcells, M.\ 1996, AJ, 111, 2238 
\bibitem{Pel97} Peletier, R.~F., 
\& Balcells, M.\ 1997, New Astronomy, 1, 349 
\bibitem{Pel99} Peletier, R.~F., 
Balcells, M., Davies, R.~L., Andredakis, Y., Vazdekis, A., Burkert, A., \& 
Prada, F.\ 1999, MNRAS, 310, 703 
\bibitem{Pru01} Prugniel, P., Maubon, 
G., \& Simien, F.\ 2001, A\&A, 366, 68 
\bibitem{Rav01} Ravindranath, S., 
Ho, L.~C., Peng, C.~Y., Filippenko, A.~V., \& Sargent, W.~L.~W.\ 2001, AJ, 
122, 653 
\bibitem{Res01} Rest, A., et al., 2001, AJ, 121, 2431
\bibitem{San04} S\'anchez-Bl\'azquez, P., 2004, Ph.D. Thesis, Universidad Complutense Madrid
\bibitem{Ter94} Terndrup, D.~M., 
Davies, R.~L., Frogel, J.~A., Depoy, D.~L., \& Wells, L.~A.\ 1994, ApJ, 
432, 518 
\bibitem{Dok95} van Dokkum, P.~G., 
\& Franx, M.\ 1995, AJ, 110, 2027 
\bibitem{Vaz96} Vazdekis, A., Casuso, 
E., Peletier, R.~F., \& Beckman, J.~E.\ 1996, ApJs, 106, 307 
\bibitem{Zoc03} Zoccali, M., et al.\ 
2003, A\&A, 399, 931 
\end{thebibliography}
\end{document}